\documentclass[12pt]{article}

\usepackage{graphics,graphicx,fullpage,natbib,multirow}
\usepackage{amsmath,amssymb,verbatim,epsfig}
\usepackage[dvipsnames,usenames]{color}
\usepackage{caption,subcaption,booktabs}

\newtheorem{defin}{\bf Definition}

\def\mul{\mbox{Mul}}

\def\geo{\mbox{Geo}}
\def\no{\mbox{N}}

\def\un{\mbox{Un}}

\def\dir{\mbox{Dir}}

\def\E{\mbox{E}}

\def\bx{{\bf x}}

\def\bF{{\bf F}}

\def\bX{{\bf X}}

\newcommand{\bpi}{\boldsymbol{\pi}}

\newcommand{\DP}{\mathcal{DP}}

\newcommand{\MP}{\mathcal{MP}}

\newcommand{\BC}{\mathcal{B}}

\newcommand{\NB}{\mathbb{N}}
\newcommand{\RB}{\mathbb{R}}

\begin{document}

\baselineskip=24pt

\title{\bf The Dirichlet process as sampling distribution}
\author{{\sc Luis E. Nieto-Barajas} \\[2mm]
{\sl Department of Statistics, ITAM Mexico} \\[2mm]
{\small {\tt luis.nieto@itam.mx}} \\}
\date{}
\maketitle

\begin{abstract}
The Dirichlet process (DP) is the most common Bayesian nonparametric prior, however, its properties as sampling distribution have not been studied nor inference on its parameters. Here we use the DP as a data generating model and make Bayesian inference on its centering measure and precision parameter. We illustrate with a sequence of histograms as observed data. In particular, we consider simulated and real datasets. 
\end{abstract}

\vspace{0.2in} \noindent {\sl Keywords}: Bayesian nonparametrics, Dirichlet process, multinomial process, multiple histograms.

\section{Introduction}
\label{sec:intro}

The Dirichlet process (DP) was introduced by \cite{ferguson:73} and is the most important model in Bayesian nonparametric statistics. Its use has been as prior distribution for the unknown sampling distribution of a data set. 

A DP process $F$ is characterised by a precision parameter $c>0$ and a centering measure $F_0$ defined on $(\Omega,\BC)$, in notation $F\sim\DP(c,F_0)$. In general, for any $K>0$ and any partition $(B_1,\ldots,B_K)$ of $\Omega$, the random vector $(F(B_1),\ldots,F(B_K))\sim\dir(cF_0(B_1),\ldots,cF_0(B_K))$, that is, has a Dirichlet distribution. 

The typical setting for the use of the DP is as follows. Consider a data set $\bX=(X_1,X_2,\ldots,X_n)$ of size $n$ that we assumed was generated from an unknown probability model $F$, i.e., $X_i\mid F \sim F$ independently. To carry out Bayesian inference, a DP prior is placed on the unknown $F$, i.e. $F\sim\DP(c,F_0)$. It is well known \citep{ferguson:74} that the posterior distribution of $F$ given the observed data $\bx$ is another DP with updated parameters $F\mid\bx\sim\DP\left(c+n,(cF_0+n\widehat{F})/(c+n)\right),$ where $\widehat{F}(\cdot)=\frac{1}{n}\sum_{i=1}^{n}I_{(-\infty,x_i)}(\cdot)$ is the empirical distribution function of the observed sample $\bx$. 

The paths of the DP process are almost surely discrete \citep{blackwell&macqueen:73}, which has been considered a drawback to model continuous data. To overcome this problem, absolutely continuous priors have been proposed by considering a continuous parametric probability model and using the DP as mixing distribution over the parameters of the model. The first to propose such DP mixtures is \cite{lo:84}. 

Although discrete, the paths of the DP are probability measures or cumulative distribution functions (CDFs). It is therefore natural to assume that, if several CDFs are available as $F_1,\ldots,F_n$, they were generated by a DP, i.e. $F_i\sim\DP(c,F_0)$ for $i=1,\ldots,n$, and make Bayesian inference on the unknown parameters of the model $c$ and $F_0$. 

Due to the advancement of technology, huge data sets are now being generated, and it becomes difficult to store and analyse them. A statistic commonly used to summarise the information is the frequency table whose graphical representation is a histogram. The manipulation and analysis of several histograms is now of interest. 

A frequency table is also known in the literature as binned data. The analysis of binned data has been dominated by the use of mixture models with normal kernels. For instance \cite{same&al:06} used an EM algorithm for dealing with the missing exact observations and made emphasis on clustering. \cite{alston&mengersen:10} dealt with the problem in a Bayesian way by extending their Gibbs sampler to sample from the unobserved data points. More recently, \cite{martinez&diaz:24} used a product partition model as the prior for the latent clusters to find groups in the data. Finally, \cite{warr&al:26} defined their mixture model hierarchically with multiple layers with Dirichlet process mixture models in one of them. 

These previous models analyse a single histogram and emphasis is on density estimation and finding groups in the binned data. For the analysis of multiple histograms, there are fewer proposals. \cite{billard&kim:17} proposed a hierarchical divisive clustering algorithm to produce groups of histograms in $\RB^p$. On the other hand, \cite{warr&al:26} also consider the clustering of multiple histograms in $\RB$. 

In this context our proposal is unique in the sense that we want to analyse multiple histograms (CDFs) in $\RB$ using the Dirichlet process as a data generating model and make inference on its parameters. 

Our goal is achieved through the following sections: In Section \ref{sec:model} we define our problem setting and notation. In Section \ref{sec:post} we carry out Bayesian inference on the model parameters and obtain posterior distributions. We illustrate the use of the model in Section \ref{sec:illust} and conclude in Section \ref{sec:disc}.

\section{Model}
\label{sec:model}

A frequency table is constructed by predefining a sequence of fixed points $\tau_0<\tau_1<\cdots<\tau_k$, with $\tau_j\in\RB$, which partition the support of the data into disjoint intervals $B_j=(\tau_{j-1},\tau_j]$, for $j=1,\ldots,k$. The relative frequency of the number of data points in the interval $B_j$ is denoted by $F(B_j)$ and the collection of all relative frequencies is denoted by $F=(F(B_1),\ldots,F(B_k))$. 

A histogram is a graphical representation of the frequency table and is usually done by plotting adjacent bars at locations $B_j$ and heights $F(B_j)/(\tau_j-\tau_{j-1})$ to define the probability density 
\begin{equation}
\label{eq:fx}
f(x)=\sum_{j=1}^k\frac{F(B_j)}{\tau_j-\tau_{j-1}}I(x\in B_j).
\end{equation}

Let us assume that our data consist of a collection of $n$ histograms $F_1,\ldots,F_n$ that we assume were generated by a DP, i.e. 
\begin{equation}
\label{eq:model}
F_i\mid c,F_0\sim\DP(c,F_0) 
\end{equation}
independently. In practice, histograms are not necessarily defined on the same partition, so we consider that each histogram $F_i$ has its own partition $\Pi_i=\{B_{i,j}\}$ with $B_{i,j}=(\tau_{i,j-1},\tau_{i,j}]$ for $j=1,\ldots,k_i$ and $k_i$ the partition size. 

In this case, the likelihood for $(c,F_0)$ is given by
$$p(F_1,\ldots,F_n\mid c,F_0)=\prod_{i=1}^n\dir\left(F_i(B_{i,1},\ldots,F_i(B_{i,k_i})\mid cF_0(B_{i,1}),\ldots,cF_0(B_{i,k_i})\right),$$
where $\dir(\bx\mid\bpi)$ denotes a Dirichlet density for $\bx$ with parameter vector $\bpi$. Substituting the form of the density, the likelihood becomes
$$p(F_1,\ldots,F_n\mid c,F_0)=\prod_{i=1}^n\Gamma(c)\prod_{j=1}^{k_i}\frac{F_i(B_{i,j})^{cF_0(B_{i,j})-1}}{\Gamma(cF_0(B_{i,j}))}I\left(\sum_{j=1}^{k_i}F_i(B_{i,j})=1\right),$$
where $\sum_{j=1}^{k_i}F_0(B_{i,j})=1$. 

To simplify the form of the likelihood, consider a common partition $\Pi^*$ constructed as the intersection of all partition elements $\{B_{i,j}\}$, in notation, $\Pi^*=\bigcap_{i=1}^n\Pi_i=\{B_j^*\}$, for $j=1,\ldots,k$ such that $B_j^*=(\tau_{j-1}^*,\tau_j^*]$ with the usual ordering $\tau_0^*<\tau_1^*<\cdots<\tau_k^*$. Furthermore, let us consider the reparameterisation $G=c F_0$ so that $G$ is an un-normalised measure. Finally, the likelihood for $G$ has the form
\begin{equation}
\label{eq:lik}
p(F_1,\ldots,F_n\mid G)=\Gamma^n(c)\prod_{j=1}^k\frac{\left\{\prod_{i=1}^n F_i(B_j^*)\right\}^{G(B_j^*)-1}}{\Gamma^n(G(B_j^*))}.
\end{equation}
Note that the common partition $\Pi^*$ is a refinement of each of the original partitions $\Pi_i$, $i=1,\ldots,n$. For $B_j^*=(\tau_{j-1}^*,\tau_j^*]\subseteq B_{i,j'}=(\tau_{i,j'-1},\tau_{i,j'}]$, we use \eqref{eq:fx} to define $F_i(B_j^*)=F_i(B_{i,j'})(\tau_j^*-\tau_{j-1}^*)/(\tau_{i,j'}-\tau_{i,j'-1})$, which corresponds to a uniform distribution of the probability in the original partition set $B_{i,j'}$. 

\section{Bayesian inference}
\label{sec:post}

We propose to carry out Bayesian inference on the reparameterised model parameters $(c,G)$ and recover the centering measure as $F_0=G/c$. Since $F_0$ is a proper probability distribution, the probability assigned to the common partition $\Pi^*$ must satisfy $\sum_{j=1}^kF_0(B_j^*)=1$, which in terms of $G$ is equivalent to $\sum_{j=1}^kG(B_j^*)=c$. Therefore, the parameter $c$ is the total mass of the un-normalised measure $G$. 

Taking into account the previous point and looking at the likelihood \eqref{eq:lik}, we propose placing a multinomial process as prior distribution for $G$ given $c$. A multinomial process \citep[e.g.][]{nieto:21} denoted as MP, is defined, similar to a DP,  through its finite dimensional distributions. For any $K>0$ and any partition $(B_1,\ldots,B_K)$ of $\Omega$, the random vector $(G(B_1),\ldots,G(B_K))\sim\mul(c,G_0(B_1),\ldots,G_0(B_K))$, that is, a multinomial distribution with number of trials $c$ and probabilities $G_0(B_j)$, $j=1,\ldots,k$. The density, using the common partition $\Pi^*$, is given by
\begin{equation}
\label{eq:mul}
p(G\mid c)=c\,!\prod_{j=1}^k\frac{G_0(B_j^*)^{G(B_j^*)}}{G(B_j^*)!}I\left(\sum_{j=1}^kG(B_j^*)=c\right).
\end{equation}

In notation, we say $G\mid c\sim\MP(c,G_0)$, with $c\in\NB$ and $G_0$ a probability measure. We complete our prior by taking a distribution for an integer valued random variable with support in the set of natural numbers, say $c\sim\geo(\pi_0)$, a geometric distribution with success probability $\pi_0$. 

We combine our prior knowledge $p(c,G)=\mul(G\mid c,G_0)\geo(c\mid\pi_0)$ with the likelihood \eqref{eq:lik} via the Bayes Theorem and obtain the posterior distribution. This is characterised through the full conditional distributions given as follows.

\begin{enumerate}
\item[(i)] Conditional posterior for $G$
$$p(G\mid c,\bF)\propto\prod_{j=1}^k\frac{\left\{G_0(B_j^*)\prod_{i=1}^nF_i(B_j^*)\right\}^{G(B_j^*)}}{\Gamma\left(G(B_j^*)+1\right)\Gamma^n\left(G(B_j^*)\right)}I\left(\sum_{j=1}^nG(B_j^*)=c\right).$$
This multivariate distribution is easily updated one element $G(B_j^*)$ at a time, for $j=1,\ldots,k-1$, from the following univariate conditional distribution
$$\hspace{-1cm}p(G(B_j^*)\mid c,\bF,G(B_l^*),l\neq j)\propto\frac{\left\{\frac{G_0(B_j^*)}{G_0(B_k^*)}\prod_{i=1}^n\frac{F_i(B_j^*)}{F_i(B_k^*)}\right\}^{G(B_j^*)}I\left\{G(B_j^*)\leq c-\sum_{l\neq j}^{k-1}G(B_l^*)\right\}}{\Gamma\left(G(B_j^*)+1\right)\Gamma^n\left(G(B_j^*)\right)\Gamma\left(G(B_k^*)+1\right)\Gamma^n\left(G(B_k^*)\right)},$$
where $G(B_k^*)=c-\sum_{l=1}^{k-1}G(B_l^*)$.
\item[(ii)] Conditional posterior for $c$
$$p(c\mid G,\bF)\propto \frac{\Gamma(c+1)\Gamma^n(c)\left\{\right(1-\pi_0)G_0(B_k^*)\prod_{i=1}^n F_i(B_k^*)\}^c}{\Gamma\left(c-\sum_{j=1}^{k-1}G(B_j^*)+1\right)\Gamma^n\left(c-\sum_{j=1}^{k-1}G(B_j^*)\right)}I\left(c\geq \sum_{j=1}^{k-1}G(B_j^*)\right).$$
\end{enumerate}

Posterior inference will therefore require the implementation of an MCMC with posterior conditional distributions (i) and (ii). Both distributions are discrete, so, in principle, sampling could be performed directly by evaluating the density for a range of values and normalising; however, since sufficient statistics are $\prod_{i=1}^n F_i(B_j^*)$ for $j=1,\ldots,k$ and each $F_i(B_j^*)\in(0,1)$, evaluating the conditional posterior densities will induce numerical problems, even for small $n$. 

An alternative solution for sampling from (i) and (ii) is to implement Metropolis-Hastings (MH) steps \citep[e.g.][]{tierney:94}, overcoming the numerical problem of computing the acceptance probability as the ratio of the conditional densities evaluated at two different values. We will follow this approach for our implementation with random walk proposals with uniform distributions and an amplitude of $\pm\delta$. We tune $\delta_G$ and $\delta_c$ appropriately to achieve acceptance rates between $30\%-40\%$. The code was implemented in Fortran and can be run from R. 

Once the posterior distributions of the model parameters $c$ and $G$ have been characterised, we can compute the posterior predictive distribution of the density \eqref{eq:fx} using the common partition $B_j^*$ and $F_0(B_j^*)$ as relative frequencies. This would allow us to estimate, point wise or via intervals, the probability for any set $B\in\RB$. 

\section{Illustration}
\label{sec:illust}

\subsection{Simulation study}

To illustrate the performance of our model and test our posterior sampling algorithm, we  first consider a simulation study. We sample data from a mixture of two normals with the following specification: 
$$f(x)=\pi\no(x\mid\mu_1,\sigma_1^2)+(1-\pi)\no(x\mid\mu_2,\sigma_2^2),$$
where $\mu_1=-2$, $\mu_2=2$, $\sigma_1=\sigma_2=1$ and $\pi=0.3$. We consider two scenarios for the study. 

The first scenario consists of taking $n\in\{10,20,30\}$ samples of size $s\in\{50,100,200\}$ and computing a frequency table for each sample. The bins are common for each sample from $-5.5$ to $5.5$ with a width of $1.5$ for the first and last bin, and $1$ for the rest. This leads to a partition of size $k=10$. Specifically, $B_1^*=(-5.5,4]$, $B_j^*=(j-6,j-5]$, for $j=2,\ldots,9$ and $B_{10}^*=(4,5.5]$. Since our model \eqref{eq:model} assumes that the relative frequencies associated with each partition element, $F_i(B_j^*)$, are Dirichlet distributed, they must be strictly positive. So, zero frequencies are not allowed. We assign a small positive value, say $0.001$, to all zero observed frequencies and renormalise for all partition elements such that $\sum_{j=1}^k F_i(B_j^*)=1$ for all $i=1,\ldots,n$. 

We implemented our model and, to specify the prior for $G\mid c$, we took two options $G_0=\un(-5.5,5.5)$ and $G_0=\no(0,4)$ truncated to the partition range. For the prior on $c$ we took $\pi_0=0.01$, so that a-priori $\E(c)=99$. The MH steps were tuned with $\delta_G=2$ and $\delta_c=3$. The MCMC was run for 550,000 iterations with a burn in of 50,000. We kept one of every 100$^{th}$ iteration to compute posterior summaries. Each run took on average 30 seconds. 

To compare, we computed the logarithm of the pseudo marginal likelihood LPML \citep{geisser&eddy:79}. These values are presented in Table \ref{tab:sim1}. We first note that the LPML values are the same or larger when $G_0$ is uniform (first row) compared to the normal (second row). The fit improves as we increase $n$ and $s$, however, the number of histograms available $n$ is not as important as having produced them with a larger number of data points $s$. In particular, the fitting with $n=20$ and $s=200$ is better than with $n=30$ and $s=50$. We took the best fit, $n=30$ and $s=200$, to produce further inferences. 

\begin{figure}
\centering
\includegraphics[scale=0.8]{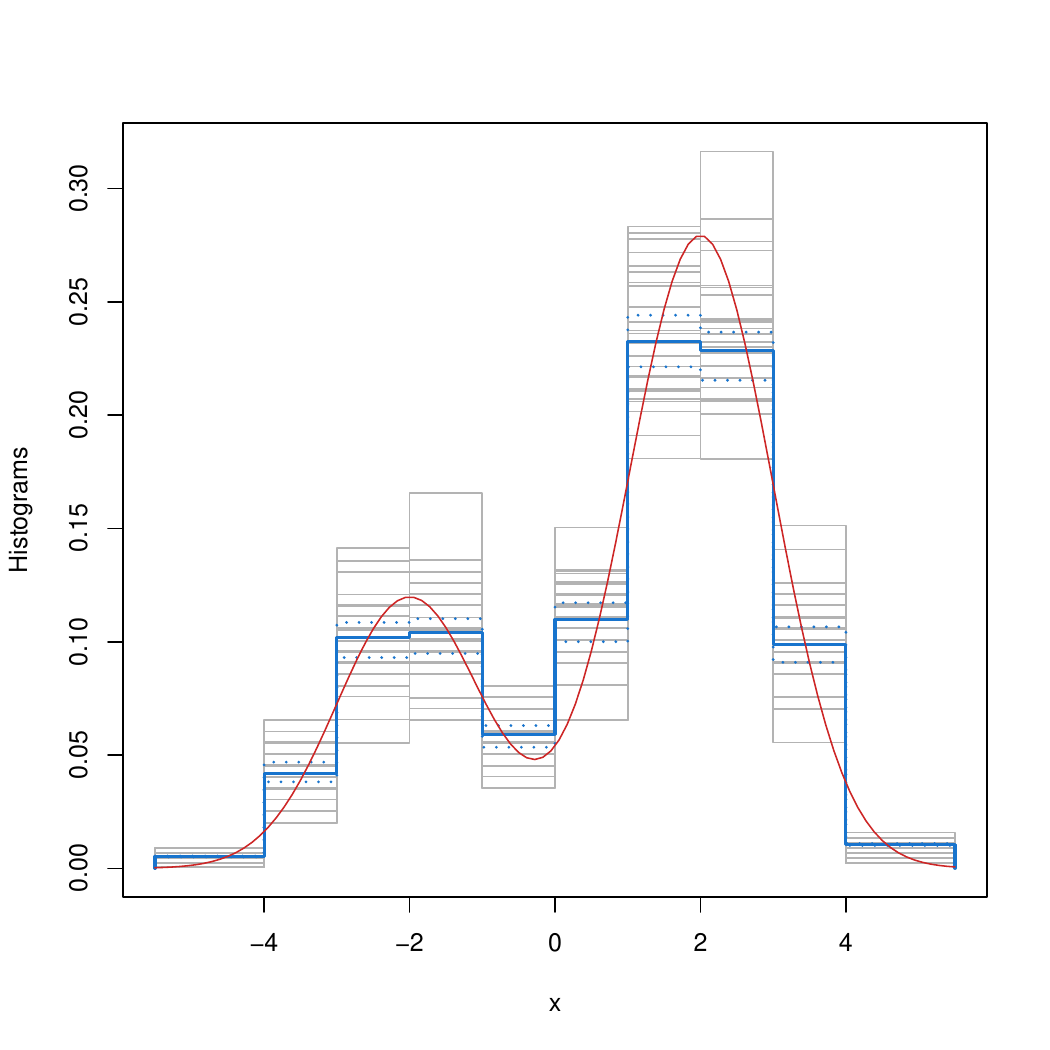}
\caption{Simulation study scenario 1. Observed histograms (grey), posterior point estimate (darker line), 95\% CI (dotted line) and real density (curved line).}
\label{fig:sim1}
\end{figure}

The posterior distribution for $c$ is concentrated around the value $127$ with a 95\% credible interval (CI) $[120,132]$. We recover $F_0=G/c$ and make posterior inferences. Figure \ref{fig:sim1} shows a graph of the data as overlapped light histograms. Our posterior point estimate is shown as a darker line together with a 95\% CI as dotted line. We also include the original density as a solid curved line. As expected, our estimate follows closely the true density but in a stepwise fashion. 

\begin{figure}
\centering
\includegraphics[scale=0.8]{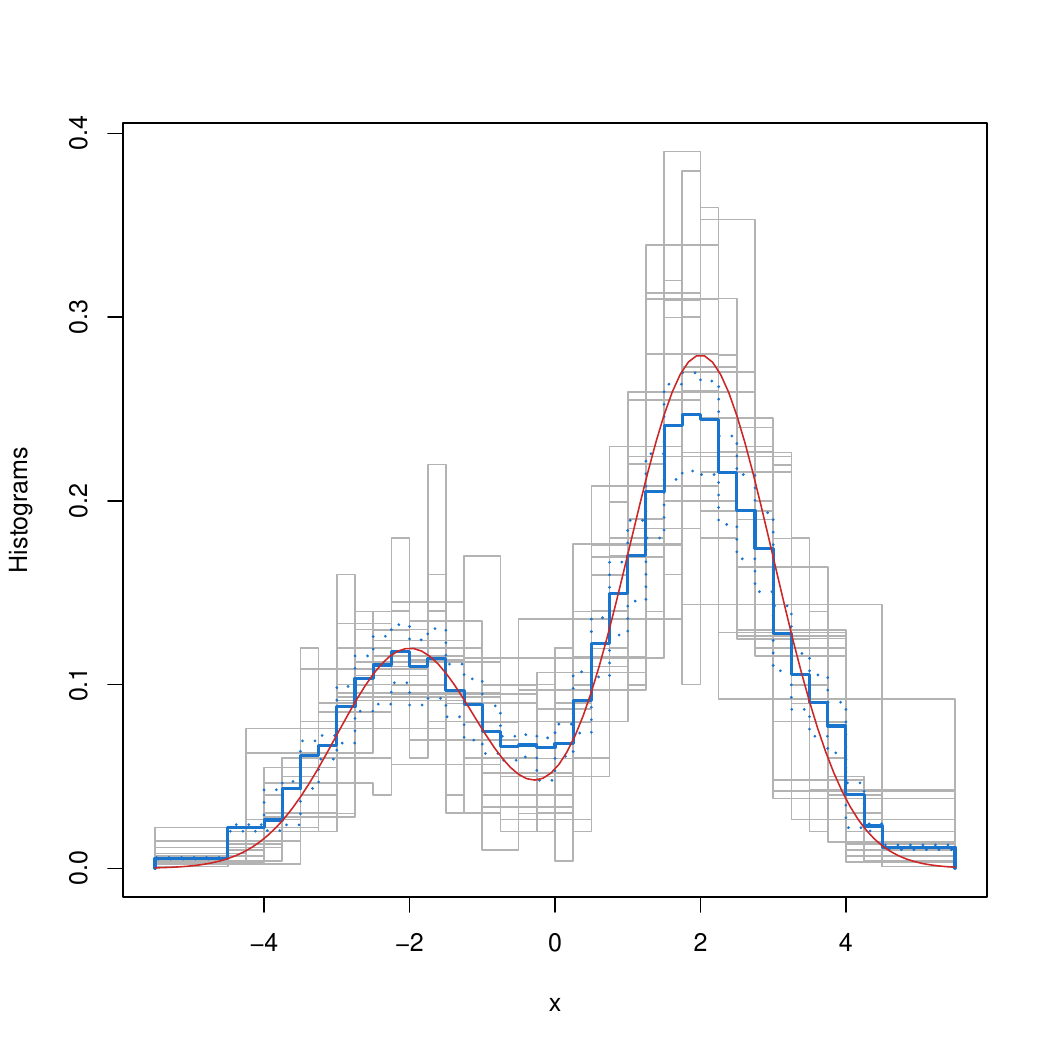}
\caption{Simulation study scenario 2. Observed histograms (grey), posterior point estimate (darker line), 95\% CI (dotted line) and real density (curved line).}
\label{fig:sim2}
\end{figure}

The second scenario consists of taking $n\in\{10,20\}$ samples of size $s=200$ and computing a frequency table for each sample. This time, the bins are random for each sample and the partition size is fixed $k_i\in\{6,8,10,12,14\}$ for $i=1,\ldots,n$. We also fixed the extremes $\tau_{i,0}=-5.5$ and $\tau_{i,k_i}=5.5$ and the remaining $\tau_{i,j}$, $j=1,\ldots,k_i-1$ were randomly selected from the set $\{-4.5,-4.25,-4.00\ldots,4.25,4.5\}$ and ordered. This results in the production of histograms that might not resemble the form of the true density. These are depicted in Figure \ref{fig:sim2} as overlapped light lines. 

We created a common partition as described in Section \ref{sec:model}. This results in having $k=38$ partition elements. Again, zero frequencies were assigned the value of $0.001$ followed by a normalisation. 

MCMC specifications were the same as in the previous scenario and the prior specifications were $G_0=\un(-5.5,5.5)$ and $\pi_0=0.01$. Each run took on average 100 seconds. LPML values are reported in Table \ref{tab:sim2}. These values increase monotonically as we increase the number of partition elements $k_i$. Additionally, having more histograms is better. We took $k_i=14$ and $n=20$ to produce posterior inferences. 

The posterior distribution for $c$ is concentrated around $181$ with a 95\% CI $[167,199]$. The posterior estimates for $F_0$ are shown in Figure \ref{fig:sim2} as darker line (posterior mean) and as dotted lines (95\% CI). Again, our estimate follows closely the path of the true density. The partial information provided by each individual histogram is appropriately summarised by the model. 

\subsection{Real data analyses}

The National Institute of Statistics and Geography in Mexico produces annual labor indicators for the 2,478 municipalities. Among the variables measured are the economically active population (EAP) and the informally occupied population (IOP). Data is available at \texttt{https://www.inegi.org.mx/programas/ilmm/\#tabulados}. 

Tabular data are reported using frequency tables for each year from 2017 to 2024, that is, there are $n=8$ available years. Histograms for EAP variable have different ranges according to the year, but in all cases bin lengths are 2 percentage points. For 2017 range goes from 32\% to 78\%, for 2018 range goes from 30\% to 74\%, for 2019 range goes from 32\% to 76\%, for 2020 range goes 42\% to 74\%, for 2021 range goes from 42\% to 76\%, for 2022 range goes from 30\% to 84\%, for 2023 range goes from 38\% to 88\%, and for 2024 range goes from 40\% to 90\%. 

We constructed a common partition of size $k=14$ with the following interval limits: $\{30,46,48,\ldots,68,70,90\}$. The first and last intervals have larger bin lengths to avoid small frequencies and having numerical problems. 
We applied our method and ran the MCMC for 550,000 iterations with a burn in of 50,000 and a thinning of 100. We took a uniform distribution as $G_0$ over the range of the common partition and for $c$ we took two options of $\pi_0\in\{0.01,0.001\}$. The LPML statistics are the same, $281$, for the two choices of $\pi_0$. 

\begin{figure}
\centering
\includegraphics[scale=0.8]{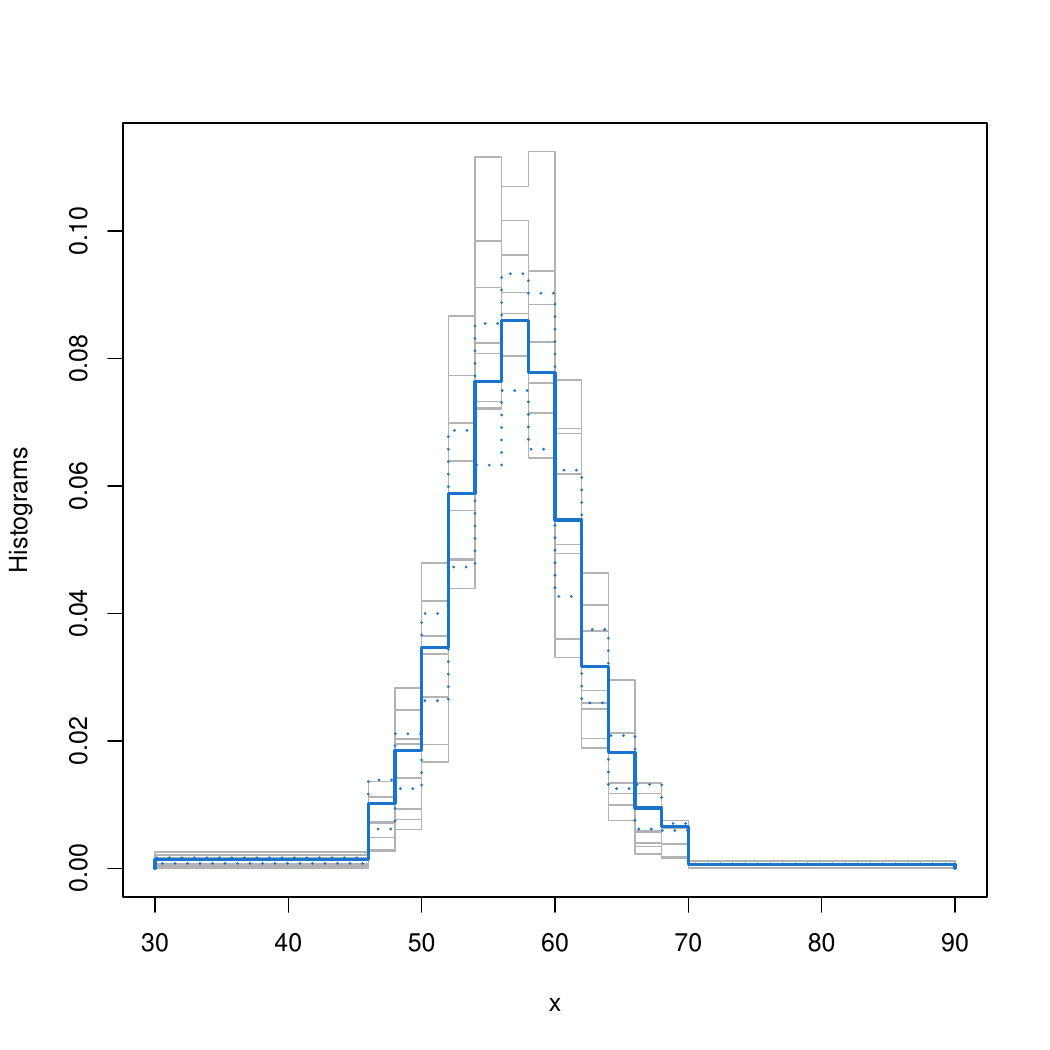}
\caption{Real data AEP. Observed histograms (grey), posterior point estimate (darker line) and 95\% CI (dotted line).}
\label{fig:pea}
\end{figure}

Posterior estimates for $F_0$ are included in Figure \ref{fig:pea} as point estimate (darker line) together with 95\% CI (dotted line). The density mode lies in the interval $(56,58]$. In terms of context, we can say that most municipalities have a percentage of around 57\% of their population economically active, with a minimum of 46\% and a maximum of 70\%. Posterior estimates for the precision parameter $c$ are 77 (posterior mean) and a 95\% CI of $[70,84]$. 

Histograms for the second variable, IOP, all have bin sizes of 5 percentage points and ranges are from 20 to 100 for the years 2017, 2020, 2022, 2023 and 2024; and from 25 to 100 for the remaining years 2018, 2019 and 2021. We created a common partition of size $k=15$ with interval limits $\{20,30,35,\ldots,100\}$. Only the first interval has a larger length of 10 percentage points. 

\begin{figure}
\centering
\includegraphics[scale=0.8]{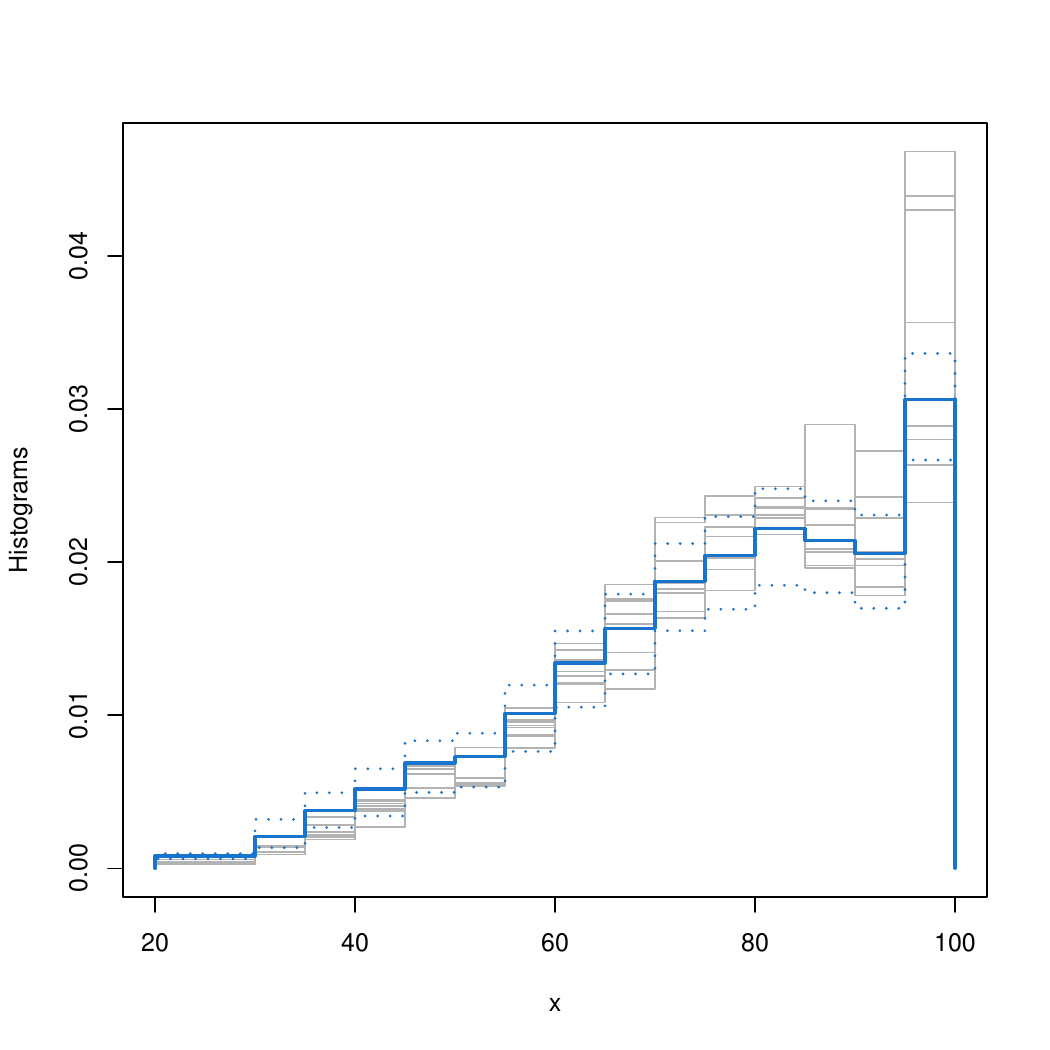}
\caption{Real data IOP. Observed histograms (grey), posterior point estimate (darker line) and 95\% CI (dotted line).}
\label{fig:iop}
\end{figure}

We fitted our model with the same MCMC and prior specifications as for the analysis of the previous variable. Again, LPML statistics are the same, $336$, for both $\pi_0$. Posterior inferences for $c$ produce a point estimate of $127$ with a 95\% CI of $[99,161]$. Posterior inferences for $F_0$ are reported in Figure \ref{fig:iop} where both data points (histograms) and density estimates are included. Our posterior mean perfectly summarises the information of the $n=8$ histograms. Interestingly, for around [13\%,17\%] of the municipalities in Mexico, with 95\% probability, close to 100\% of their occupied populations is informal. This would be important to consider for tax revenue purposes.

\section{Discussion}
\label{sec:disc}

We have effectively used the Dirichlet process as a sampling distribution for a collection of histograms. We resorted to a common partition to pull strength from the different histograms. The parameters of the Dirichlet process are two, a non-negative parameter $c$ and a probability measure $F_0$. 

A well known model for clustering observations is the hierarchical Dirichlet process \citep{teh&al:06} which consists of assuming a Dirichlet process as Bayesian nonparametric prior for data such that the centering measure $F_0$ is itself another Dirichlet process. Following this setting, we could have assumed a Dirichlet process as a prior distribution for our $F_0$. Instead, we proposed to use a multinomial process as prior distribution for the un-normalised measure $G=cF_0$. The reason for doing so is due to the form of the likelihood \eqref{eq:lik} that partially conjugates with the multinomial process. 

Implementing our model is quite straightforward; however, there are numerical problems that the user must be aware of. First, the sampling model does not allow data to have zero relative frequencies in a partition interval, they all have to be strictly positive. Second, very small relative frequencies also create numerical problems. We advise joining some adjacent partition intervals and adding the relative frequencies. Once the numerical issues are resolved, the MCMC runs pretty fast. The examples considered here were run on a machine with an Intel Xenon processor at 3.00 GHz with 24GB of RAM and a Linux operating system. They took less than 2 minutes to run. 

Future work could be to consider covariates or time to better model the data. On the other hand, we could explore the possibility to use other almost surely discrete nonparametric models, such as stick-breaking or normalised measures.

\section*{Acknowledgements}

The author acknowledges support from \textit{Asociaci\'on Mexicana de Cultura, A.C.} and ICERM-Brown University for the invitation to the workshop on ``Nonparametric Bayesian Inference, Computational Issues'' where this work was motivated on. 


\bibliographystyle{natbib}

\begin{thebibliography}{99}

\bibitem[Alston and Mengersen, 2010]{alston&mengersen:10} 
Alston, C.L. and Mengersen, K.L. (2010). Allowing for the effect of data binning in a Bayesian normal mixture model. \textit{Computational Statistics and Data Analysis} \textbf{54}, 916--923.

\bibitem[Billard and Kim, 2017]{billard&kim:17} 
Billard, L. and Kim, J. (2017). Hierarchical clustering for histogram data. \textit{Wiley Interdisciplinary Reviews: Computational Statistics} \textbf{9}, e1405.

\bibitem[Blackwell and MacQueen, 1973]{blackwell&macqueen:73} 
Blackwell, D. and MacQueen, J.B. (1973). Ferguson distributions via p\'olya urn schemes. \textit{Annals of Statistics} \textbf{1}, 353-355.

\bibitem[Ferguson, 1973]{ferguson:73}
Ferguson, T.S. (1973). A Bayesian analysis of some nonparametric problems. \textit{Annals of Statistics} \textbf{1}, 209--230.

\bibitem[Ferguson, 1974]{ferguson:74}
Ferguson, T.S. (1974). Prior distributions on spaces of probability measures.\textit{Annals of Statistics} \textbf{2}, 615--629.

\bibitem[Geisser and Eddy, 1979]{geisser&eddy:79}
Geisser, S. and Eddy, W.F. (1979). A predictive approach to model selection.\textit{Journal of the American Statistical Association} \textbf{74}, 153--160.

\bibitem[Lo, 1984]{lo:84}
Lo, A.Y. (1984). On a class of Bayesian nonparametric estimates. I. Density estimates. \textit{Annals of Statistics} \textbf{12}, 351--357.

\bibitem[Mart\'inez and D\'iaz-Avalos, 2024]{martinez&diaz:24}
Mart\'inez, A.F. and D\'iaz-Avalos, C. (2024). A model-based approach for clustering binned data. Preprint. arXiv: 2409.07738.

\bibitem[Nieto-Barajas, 2021]{nieto:21}
Nieto-Barajas, L.E. (2020). A class of dependent dirichlet processes via latent
multinomial processes. \textit{Statistics} \textbf{55}, 1169--1179.

\bibitem[Sam\'e et al., 2006]{same&al:06}
Sam\'e, A., Ambroise, C. and Govaert, G. (2006). A classification EM algorithm for binned data. \textit{Computational Statistics and Data Analysis} \textbf{51}, 466--480.

\bibitem[Teh et al., 2006]{teh&al:06}
Teh, Y.W., Jordan, M.I., Beal, M.J. and Blei, D.M. (2006). Hierarchical dirichlet processes. \textit{Journal of the American Statistical Association} \textbf{101}, 1566--1581.

\bibitem[Tierney, 1994]{tierney:94}
Tierney, L. (1994). Markov chains for exploring posterior distributions. \textit{Annals of Statistics} \textbf{22}, 1701--1762.

\bibitem[Warr et al., 2026]{warr&al:26}
Warr, R.L., Quintana, F.A., Guglielmi, A. and Beraha, M. (2026). Bayesian mixture models for histograms: with applications to large datasets. Preprint. arXiv: 2606.24001.

\end{thebibliography}

\begin{table}[h]
$$\begin{array}{ccc|ccc|ccc} \hline \hline \\[-4mm]
\multicolumn{3}{c|}{n=10} & \multicolumn{3}{c|}{n=20} & \multicolumn{3}{c}{n=30} \\ \hline
s=50 & s=100 & s=200 & s=50 & s=100 & s=200 & s=50 & s=100 & s=200 \\
\hline
169 & 222 & 244 & 281 & 445 & 499 & 345 & 662 & 761 \\
149 & 224 & 237 & 264 & 447 & 496 & 347 & 614 & 749 \\
\hline \hline
\end{array}$$
\caption{LPML statistics for first scenario of the simulation study. First row $G_0=U(-5.5,5.5)$, second row $G_0=\no(0,4)$.}
\label{tab:sim1}
\end{table}

\begin{table}[h]
$$\begin{array}{ccccc} \hline \hline \\[-4mm]
k_i=6 & k_i=8 & k_i=10 & k_i=12 & k_i=14 \\ \hline
\multicolumn{5}{c}{n=10} \\ \hline
1114 & 1165 & 1166 & 1224 & 1240 \\ \hline
\multicolumn{5}{c}{n=20} \\ \hline
2258 & 2395 & 2389 & 2439 & 2446 \\
\hline \hline
\end{array}$$
\caption{LPML statistics for second scenario of the simulation study.}
\label{tab:sim2}
\end{table}

\end{document}